\documentclass[runningheads]{llncs}

\usepackage[T1]{fontenc}

\usepackage{graphicx,verbatim}

\usepackage{amsmath,amssymb}

\usepackage{booktabs}

\usepackage{multirow}

\usepackage{xcolor}

\usepackage{placeins}  
\usepackage{float}     

\begin{document}
\title{MUX-USCT: A Noise-Robust Neural Network for\\Ultrasound Computed Tomography}
\titlerunning{MUX-USCT}
%
\author{Yuchen Yuan\inst{1} \and
Hanhan Wu\inst{1} \and
Jinyang Li\inst{1} \and
Hanchen Wang\inst{2} \and
Yixuan Wu\inst{3} \and
Youzuo Lin\inst{2} \and
Lei Yang\inst{1}\thanks{Corresponding author.}}
\authorrunning{Y. Yuan et al.}
\institute{George Mason University, Fairfax, VA, USA\\
\email{\{yyuan21, hwu28, jli56, lyang29\}@gmu.edu} \and
The University of North Carolina at Chapel Hill, Chapel Hill, NC, USA\\
\email{\{hcwang, yzlin\}@unc.edu} \and
Johns Hopkins University, Baltimore, MD, USA\\
\email{yixuan\_wu@jhu.edu}}

      \maketitle              


\begin{abstract}

Deep neural networks (DNNs) have shown strong potential for ultrasound computed tomography (USCT) reconstruction in ideal noise-free environments, yet existing DNNs are vulnerable to the noisy conditions in clinical practice, as they equally treat inputs that suffer mild, moderate, or severe noise.
More challenging, the distributions of noise shift along with the environment, indicating the less effectiveness of noise-aware training, which injects a specific noise distribution into the training data.
We rethink these challenges and observe that the DNN models can become more robust to noise if we know the noise sources and filter them out.
This filtering operation is very alike the Multiplexers (or MUX), a fundamental combinational circuit in digital logic design.
However, the challenge here is that noise can happen randomly during inference; as a result, the manually predefined MUX cannot work.
To address these challenges, we propose MUX-USCT, a novel encoder-decoder DNN architecture that encodes the known acoustic acquisition geometry with an ``adaptive MUX'' that can automatically identify and filter noise, where the attention mechanism is applied in reconstructing the speed-of-sound map.
On the OpenPros benchmark, MUX-USCT reaches 6.88\,m/s MAE with 17\% fewer parameters than the leading baseline with 7.65\,m/s of MAE. Under simulated clinical noise, it remains stable across diverse degradation types that cause geometry-agnostic baselines to fail.
Results show that the attention distributions in MUX-USCT provide interpretable indicators of the signal quality between pairs of transducers.

\keywords{USCT \and Robustness \and DNN \and MUX \and Attention.}

\end{abstract}

\section{Introduction}

Ultrasound computed tomography (USCT) reconstructs quantitative tissue maps from 
acoustic waveform data, offering a radiation-free modality for breast screening~\cite{ref_breast_usct,sheng2024aps,vargas2025breast,zhou2024coupling}, prostate imaging~\cite{mohsen2022role,ref_openpros} and many others~\cite{fradi2022real,hatamizadeh2022gradvit,Survey-2025-lin,long2023deep,lozenski2024learned,ref_martensson,pfistner2024improved,xie2024usct}. Unlike conventional B-mode ultrasound, USCT exploits the full acoustic wavefield to achieve quantitative accuracy~\cite{ruiter2023ultrasound}, but the underlying inverse problem, full waveform inversion (FWI)~\cite{ref_fwi}, is computationally demanding. Deep learning, in particular DNN-based approaches, has shown strong potential for data-driven USCT reconstruction: InversionNet~\cite{ref_inversionnet}, a fully convolutional encoder-decoder, and ViT-Inversion~\cite{ref_vit,hatamizadeh2022gradvit,ref_openpros}, a Vision Transformer variant, both achieve strong accuracy on clean simulated data. Yet these DNNs are vulnerable to the noisy
conditions in clinical practice, as they treat the input waveform tensor as a generic feature map, which cannot distinguish noisy data~\cite{lozenski2024learned,zhang2020data}.



\begin{figure}[t]
\centering
\includegraphics[width=\textwidth]{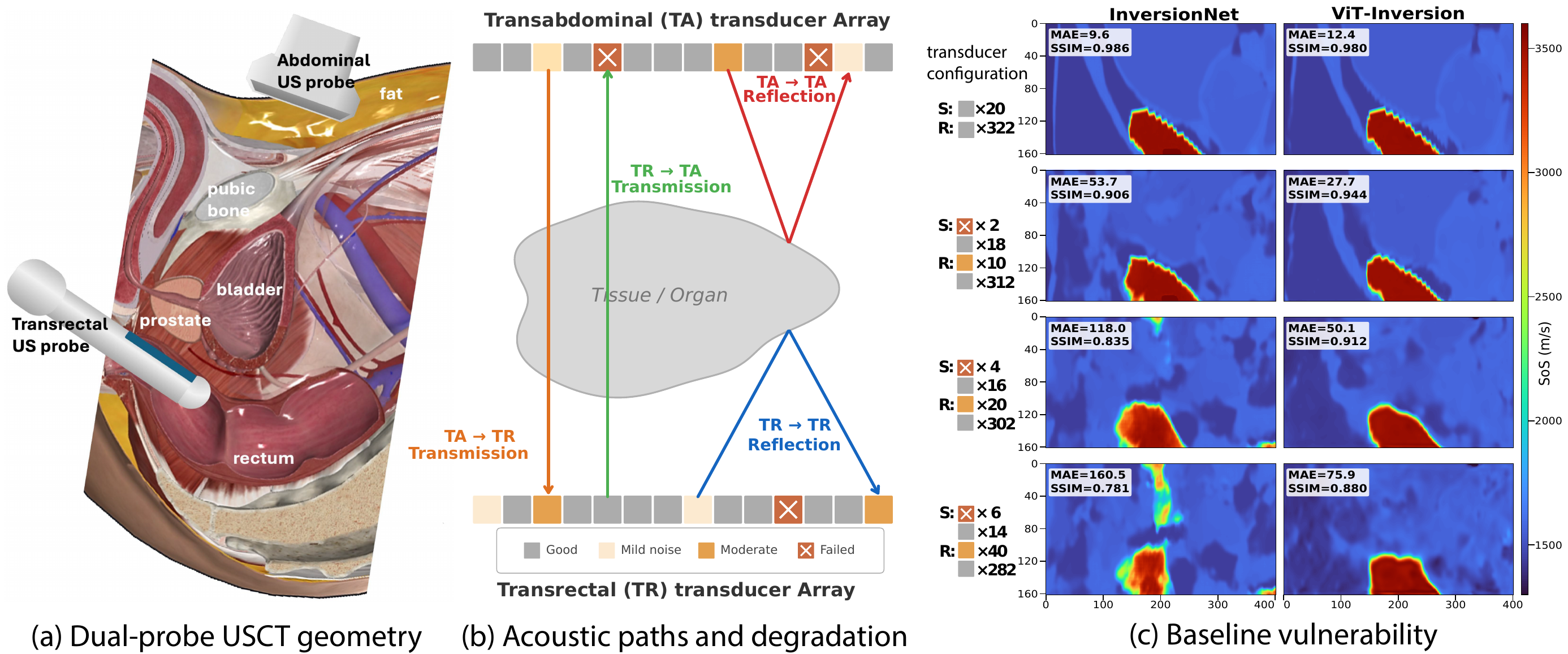}
\caption{(a)~Dual-probe prostate USCT geometry~\cite{ref_openpros}. (b)~Four acoustic paths defined by source and receiver arrays, with transducer degradation types. (c)~Reconstruction quality under progressively severe degradation with models trained on clean data.}
\label{fig:intro_motivation}
\end{figure}

On the other hand, the hardware degrades and environmental noise commonly exists in practice:
Surveys \cite{ref_dudley_qa,ref_martensson} report that 27--40\% of probes in routine hospital use carry at least one fault.
In addition, clinical environments introduce transducer noise; for example, poor probe-tissue contact attenuates signal amplitude (a.k.a., coupling loss), which can lead to different degradation levels (mild, moderate, and failed in Fig. \ref{fig:intro_motivation}b).
Moreover, clinical noise distributions shift across sites and equipment, so noise-aware training that injects a fixed degradation profile offers limited generalization.
These practical insights reveal the need for a DNN model that can be trained on clean data while adapting to noise during inference.
Ideally, the data-driven approach will be more powerful if the results of DNN inference can help the USCT specialist locate hardware degradation and provide guidance on action in the field.



Unfortunately, the existing DNNs lack these capabilities: First, as shown in Fig.~\ref{fig:intro_motivation}c, along with the increase in noise (from 1 failed source and 5 moderately noisy receivers to 3 and 20 of each), reconstruction results using the SOTA DNNs (InversionNet~\cite{ref_inversionnet} and ViT~\cite{ref_vit}) trained on clean data will collapse. Second, these DNN models equally treat all input channels regardless of quality; degradation in any single transducer propagates unchecked through the DNN, and the resulting reconstruction offers no clue as to \emph{which} transducer is at fault.

\begin{figure}[t]
\centering
\includegraphics[width=\textwidth]{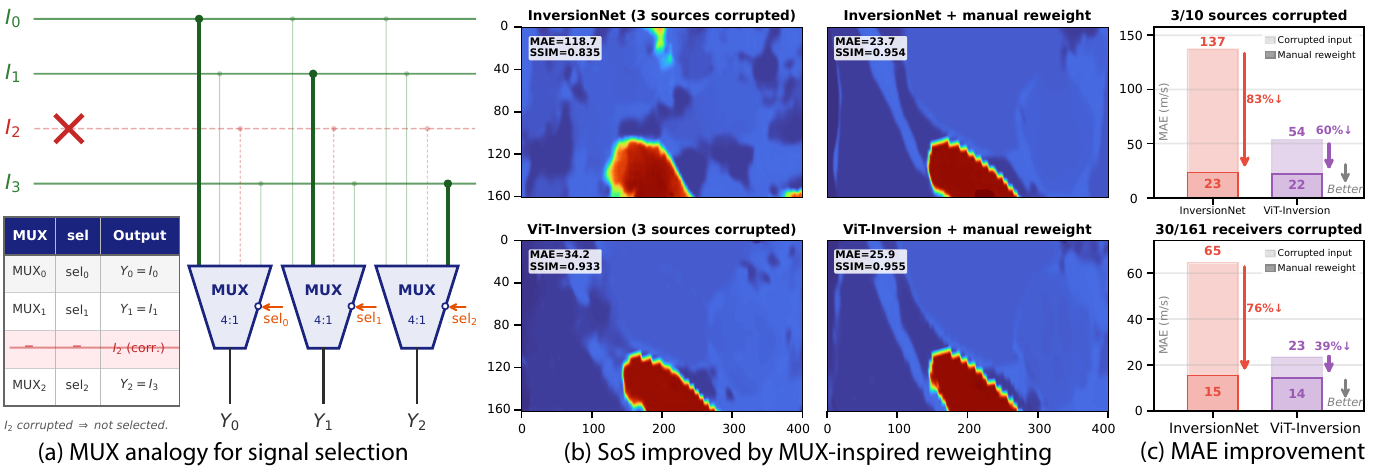}
\caption{Motivation of MUX-inspired design. (a)~Illustration of an MUX in selecting clean and corrupted inputs. (b) under 3 out of 10 source corruption, SoS improved by MUX-inspired reweighting. (c)~MAE improvements by MUX-inspired manual reweight.}
\label{fig:reweight_motivation}
\end{figure}

To achieve our goal, we first investigate the data composition and observe four different data paths in USCT with different propagation characteristics, as shown in Fig.~\ref{fig:intro_motivation}b.
By decoupling the data into paths, different paths can be weighted to reconstruct organs (e.g., Bone and Prostate).
In addition, inspired by the MUX, a fundamental combinational circuit in digital logic, given the waveform with known noise on sources, we manually filter out the noisy paths as the MUX in Fig. \ref{fig:reweight_motivation}a.
We observe a significant performance boost on both visualization and quantitative results in Fig. \ref{fig:reweight_motivation}b-c.
However, in practice, it remains challenging because noise paths are generated at runtime and cannot be identified manually using a preset configuration.

In addressing this practical challenge, we propose MUX-USCT, a novel DNN architecture.
In MUX-USCT, an encoder acts as an \emph{adaptive learned MUX} that uses an attention mechanism to assess each transducer by its signal content, which can automatically identify and filter out the corrupted transducers and compress multi-path waveforms into a compact representation.
Then, a decoder reverses this process, selectively routing the most relevant path information to each output pixel. 
The entire architecture is parametric in the number of paths and transducers, which can be set according to the acquisition geometry.






Our contributions include (1)~MUX-USCT, a novel DNN architecture with an encoder to map physical acquisition geometry into USCT reconstruction and a position-aware decoder to enable spatially selective path routing; (2)~stable robustness under diverse clinically motivated degradations with no noise-augmented training, while matching or improving clean-data accuracy at 17\% fewer parameters; (3)~interpretable attention distributions revealing transducer quality estimates and tissue-dependent spatial routing, which can assist USCT specialists in troubleshooting in the field.


\section{Method}

\textbf{Problem Formulation.}
Given a sensing system with a known acquisition geometry, the input can be described as a waveform tensor $\mathbf{X} \in \mathbb{R}^{P \times S \times T \times R}$, where $P$ acoustic paths are defined by the source--receiver probe pairing (e.g., same-probe reflection vs.\ cross-probe transmission), each path contains $S$ transducer elements that fire sequentially, and each firing is recorded over $T$ time steps at $R$ receivers. Every measurement, therefore, carries two structural tags: \emph{which path} it belongs to (encoding propagation geometry) and \emph{which element} produced it (encoding source identity).

\textbf{Design Philosophy.}
Unlike conventional approaches that ignore the structural information by flattening all channels into a single feature map, MUX-USCT retains and exploits it.
In an acoustic system, different elements and paths carry unequal information about different spatial regions; a principled reconstruction model should reflect this structure rather than treating all channels identically. MUX-USCT therefore preserves path and element identity throughout the network, allowing attention mechanisms to learn signal relevance in a data-driven manner. Because the model captures the physics of the acquisition, robustness to hardware degradation emerges as a natural byproduct: attention trained on clean data learns to weight channels by their informativeness, thereby inherently downweighting corrupted signals. Thus, even if the entire model is trained on \emph{clean data only}, requiring no noise-specific augmentation or retraining, MUX-USCT can filter out or downweight corrupted signals at inference.

\begin{figure}[t]
\centering
\includegraphics[width=0.99\textwidth]{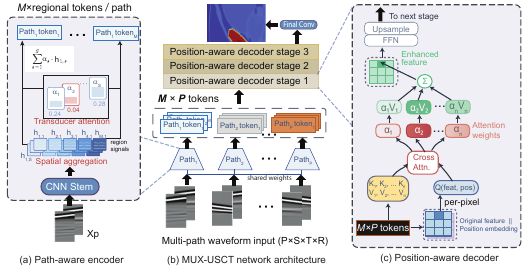}
\caption{MUX-USCT architecture: (a) details of MUX-USCT encoder; (b) MUX-USCT overall network architecture; (c) details of MUX-USCT  decoder.}

\label{fig:architecture}
\end{figure}

\textbf{Architecture Overview.}
Fig. \ref{fig:architecture}b shows the MUX-USCT overall network architecture. It contains two key components: (i)~a \emph{path-aware encoder} (see Fig. \ref{fig:architecture}a) compresses each path into $M$ regional tokens (i.e., $M\times P$ tokens in total); and (ii)~a \emph{position-aware decoder} (see Fig. \ref{fig:architecture}c) takes the output tokens of the encoder and reconstructs a 2-D tissue-property SoS map via per-pixel cross-attention.

\textbf{Path-Aware Encoder.}
Each path $\mathbf{X}_p \!\in\! \mathbb{R}^{S \times T \times R}$ is independently normalized and processed by a shared convolutional backbone (CNN Stem in Fig. \ref{fig:architecture}a), producing features $\mathbf{h}_{p,s} \!\in\! \mathbb{R}^{d}$. A \emph{transducer attention} mechanism aggregates across the $S$ elements via a learned global query $\mathbf{q} \!\in\! \mathbb{R}^{d}$:
\begin{equation}
\alpha_{p,s} \;=\; \frac{\exp\!\bigl(\mathbf{q}^\top \mathbf{k}_{p,s} / \sqrt{d}\bigr)}
     {\sum_{s'=1}^{S} \exp\!\bigl(\mathbf{q}^\top \mathbf{k}_{p,s'} / \sqrt{d}\bigr)},
\qquad
\bar{\mathbf{h}}_p \;=\; \sum_{s=1}^{S} \alpha_{p,s}\, \mathbf{v}_{p,s},
\label{eq:source_attn}
\end{equation}
where $\mathbf{k}_{p,s} {=} W_k \mathbf{h}_{p,s}$ and $\mathbf{v}_{p,s} {=} W_v \mathbf{h}_{p,s}$. The weights $\alpha_{p,s}$ act as \emph{learned select signals} analogous to MUX control lines. Since the model is trained only on clean data, the query--key inner product space reflects the statistics of intact signals; at inference, a corrupted element $s^*$ produces a key $\mathbf{k}_{p,s^*}$ that falls outside this learned subspace, receiving a low score. Softmax then concentrates mass on clean elements, filtering the corrupted channel without an explicit noise detector. Temporal attention pooling along the time axis further compresses each path into $M$ regional tokens, each summarizing a spatially distinct sub-region of that path's acoustic coverage, yielding $\mathbf{Z} \!\in\! \mathbb{R}^{M\times P \times d}$. A multi-head self-attention layer over all $M\times P$ tokens then enables information exchange across paths.

\textbf{Position-Aware Decoder.}
The decoder reverses the MUX operation: it routes the compressed $M\times P$ tokens back to spatial locations via per-pixel cross-attention~\cite{ref_detr}. At position $(i,j)$, a learned query $\mathbf{q}_{ij}$ attends to the full token set:
\begin{equation}
\mathbf{o}_{ij} \;=\; \sum_{n=1}^{M\times P} \beta_{ij,n}\, W_v' \mathbf{z}_n,
\qquad
\beta_{ij,n} \;=\; \mathrm{softmax}_n\!\Bigl(\mathbf{q}_{ij}^\top W_k' \mathbf{z}_n / \sqrt{d}\Bigr).
\label{eq:decoder_xattn}
\end{equation}
Since acoustic coverage is spatially non-uniform, each location learns to weight different path-region tokens according to its geometry (Fig.~\ref{fig:physics}c). The resulting maps $\beta \!\in\! \mathbb{R}^{H \times W \times M\times P}$, reshapable to $[H,W,P,M]$, provide interpretable per-path contribution at every pixel. A TokenToGrid module~\cite{ref_detr} first materializes a coarse spatial feature map from the tokens via this cross-attention; subsequent upsampling stages then refine it into the final tissue-property map.


\section{Experiments}

\begin{table}[b]
\centering
\caption{Simulated degradation scenarios.}
\label{tab:noise}
\setlength{\tabcolsep}{1pt}
\begin{tabular}{@{}clll@{}}
\toprule
ID & Degradation & Simulation & Clinical Origin \\
\midrule
S1 & Source noise & $-$10\,dB Gaussian per element & Faulty transducer element \\
S2 & Receiver noise & Global AWGN, 5--50\,dB SNR & Electronic/thermal interference \\
S3 & Coupling loss & 10--50\% amplitude attenuation & Poor probe-tissue contact \\
S4 & Combined & S1-S3 at mild/moderate/severe levels & Multi-fault co-occurrence \\
\bottomrule
\end{tabular}
\end{table}

\textbf{3.1 Settings.}
To evaluate the proposed MUX-USCT, we use the OpenPros prostate USCT benchmark~\cite{ref_openpros}, which includes 1000 test samples based on Finite Difference Time Domain (FDTD) simulated waveforms.
The acquisition yields $P{=}4$ acoustic paths, $S{=}10$ sources per path, $T{=}1000$ time steps, and $R{=}161$ receivers (Fig.~\ref{fig:intro_motivation}a--b). We set $M{=}5$ regional tokens per path ($P\cdot M{=}20$ total) and embedding dimension $d{=}384$; the output is a $401 \times 161$ speed-of-sound (SoS) map. 
According to different clinical failure modes~\cite{ref_martensson,ref_dudley_qa}, we created 4 simulated degradation scenarios, denoted by S1 to S4, summarized in Tab.~\ref{tab:noise}.

For comparison, MUX-USCT has three competitors, including InversionNet, ViT-Inversion, and Inv-FT.
InversionNet and ViT-Inversion are obtained from official pretrained checkpoints in \cite{ref_openpros}, which are trained on clean data.
Inv-FT applies noise-aware training \cite{ref_noise_aug}, using InversionNet with additional fine-tuning for 40 epochs on data with noise scenario S1, where 1--5 randomly selected sources per sample are corrupted with additive Gaussian noise (${\sim}$0\,dB SNR per channel).



\noindent\textbf{3.2 Robustness Evaluation.}
Table~\ref{tab:clean} shows quantitative results comparison among MUX-USCT and competitors on clean data and noisy scenario S4.
For noise-agnostic DNNs, InversionNet and ViT-Inversion, performance degrades significantly under noise S4, compared with clean data, with SSIM decreasing from 0.9893 to 0.8908.
For Inv-FT, as it is finetuned using S1, it also has the same performance degradation on S4.
Although MUX-USCT also has performance degradation on S4, its performance on S4 is close to the performance of ViT-Inversion on clean data.
What's more, MUX-USCT has 17\% fewer parameters than InversionNet, and it even achieves performance improvement on clean data.
These results validate the robustness of MUX-USCT in noisy environments and reveal that exploiting the acquisition structure can improve reconstruction.


\begin{table}[t]
\centering
\caption{Reconstruction quality comparison (best in \textbf{bold}). Moderate level is for S4.}
\label{tab:clean}
\setlength{\tabcolsep}{5pt}
\begin{tabular}{@{}cccccccc@{}}
\toprule
\multirow{2}{*}{Model} & \multirow{2}{*}{Params}  & \multicolumn{3}{c}{Clean} & \multicolumn{3}{c}{S4: Combined} \\
\cmidrule(lr){3-5} \cmidrule(lr){6-8}
 & & MAE\,$\downarrow$ & SSIM\,$\uparrow$ & PCC\,$\uparrow$ & MAE\,$\downarrow$ & SSIM\,$\uparrow$ & PCC\,$\uparrow$ \\
\midrule
InversionNet & 20.45M & 7.65 & 0.9893 & 0.9908 & 62.94 & 0.8908 & 0.9208 \\
Inv-FT & 20.45M & 12.18 & 0.9806 & 0.9766 & 69.78 & 0.9115 & 0.9353 \\
ViT-Inversion & 28.33M & 14.27 & 0.9792 & 0.9760 & 65.57 & 0.8988 & 0.8995 \\
MUX-USCT & \textbf{16.92M} & \textbf{6.88} & \textbf{0.9912} & \textbf{0.9923} & \textbf{13.49} & \textbf{0.9761} & \textbf{0.9635} \\
\bottomrule
\end{tabular}
\end{table}

\begin{figure}[t]
\centering
\includegraphics[width=\textwidth]{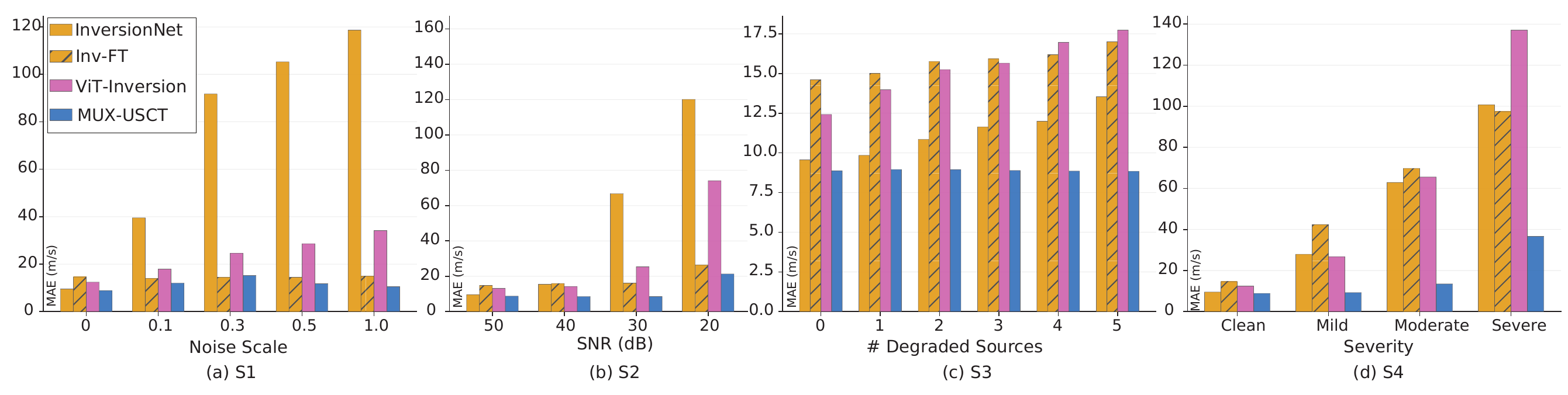}
\caption{Robustness across four clinical noise scenarios (S1--S4) in Tab. \ref{tab:noise}.}
\label{fig:clinical_noise}
\end{figure}

The robustness of MUX-USCT can be further verified on varied degradation scenarios.
Fig.~\ref{fig:clinical_noise} reports the results across all four scenarios (S1--S4) with different noise levels.
Results again show that the noise-agnostic models (InversionNet and ViT-Inversion) suffer significant performance degradation under noise during inference.
Such degradation also appears on Inv-FT for S2--S4; however, Inv-FT shows strong robustness in S1. This is because the noisy training data follow the distribution of S1.
These results reveal the limits of noise-aware training, which only works when the noise distribution is known; however, in clinical practice, environmental noise is stochastic, rendering noise-aware training inefficient.

\begin{figure}[t]
\centering
\includegraphics[width=\textwidth]{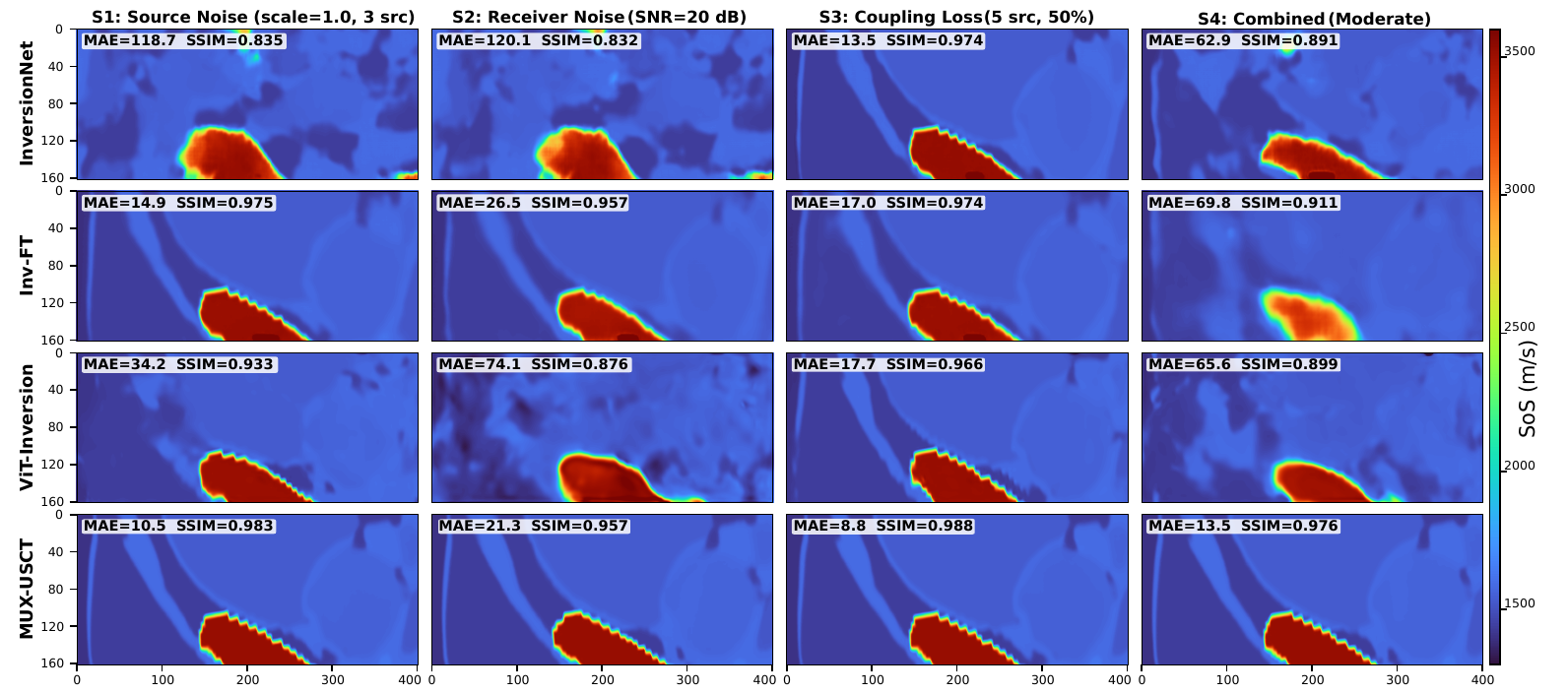}
\caption{SoS map visualization results of four models on noise scenarios (S1--S4).}
\label{fig:qualitative}
\end{figure}

On the other hand, results in Fig.~\ref{fig:clinical_noise} clearly show that MUX-USCT is robust to different scenarios.
We have an interesting observation for ``S3: Coupling loss'' in Fig.~\ref{fig:clinical_noise}c, where MUX-USCT's MAE is nearly unchanged while all baselines rise along with the increase of degraded sources.
Such robustness can also be reflected from the visualization results in Fig.~\ref{fig:qualitative}, where the last row is obtained by MUX-USCT for all scenarios, which preserves tissue boundaries.
For comparison, InversionNet and ViT-Inversion produce severe artifacts under source and receiver noise (S1, S2). Inv-FT recovers well under source noise (S1) but degrades sharply under combined degradation (S4).
These results show that MUX-USCT enables training on clean data while achieving high performance across varying levels of noise during inference.




\begin{figure}[t]
\centering
\includegraphics[width=0.92\textwidth]{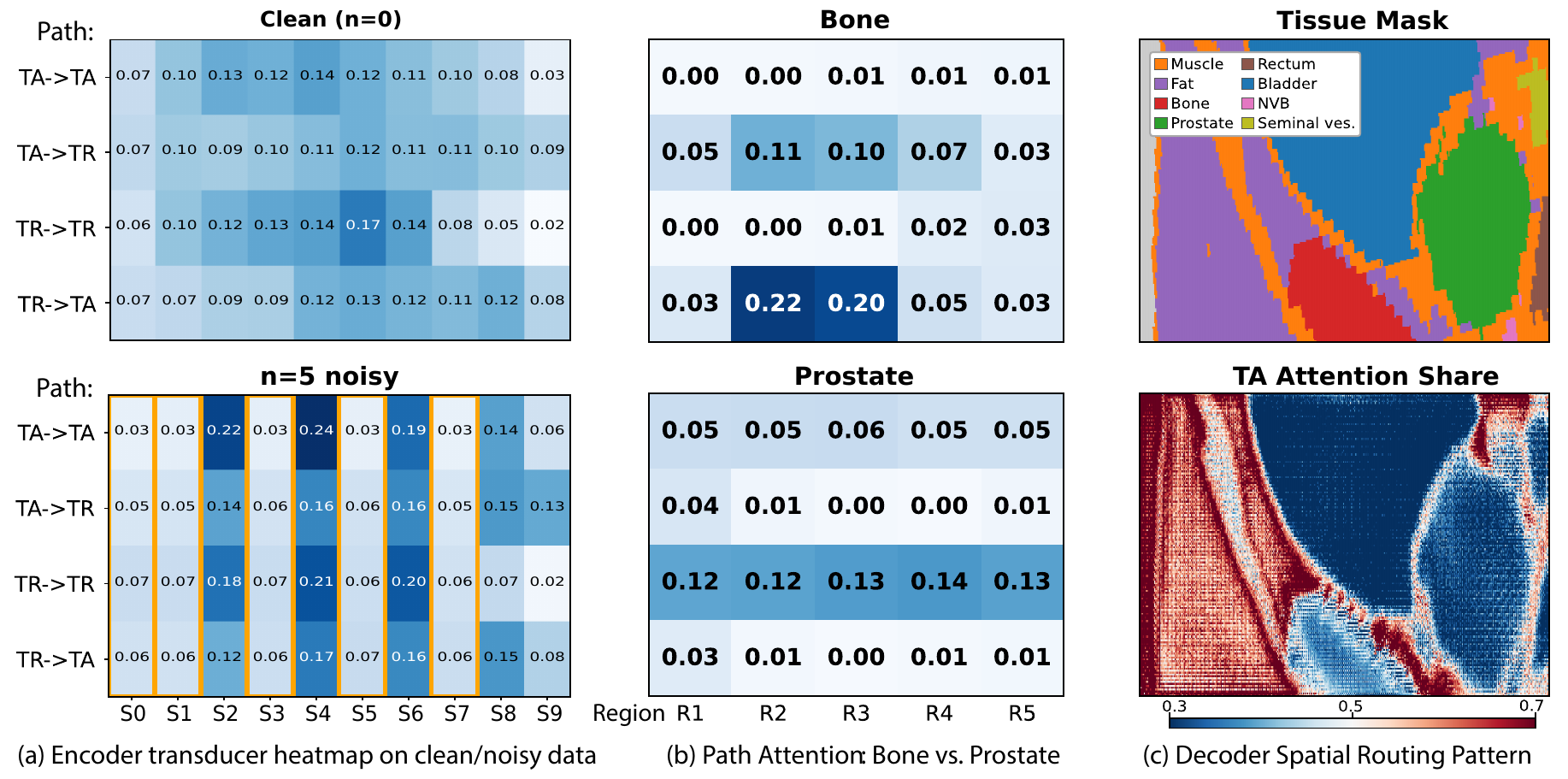}
\caption{Attention analysis: (a)~Encoder transducer attention on clean vs.\ corrupted data; (b)~Per-tissue path attention for bone vs.\ prostate; (c)~Decoder spatial routing: TA attention share varies smoothly across the image.}
\label{fig:physics}
\end{figure}

\noindent\textbf{3.3 Physics Accordance and  Clinical Guidance.}
To fundamentally reveal why MUX-USCT works, we record the encoder's transducer attention.
Fig.~\ref{fig:physics}a shows the value of attention weights on clean data and data with 5 corrupted sources. 
From the figure, we use orange boxes to highlight the attention with a small value. Interestingly, these small attention weights correspond to 5 corrupted sources, indicating these data are filtered out.
In clinical monitoring, we can use these attention weights as health indicators of transducers.

Fig.~\ref{fig:physics}b reveals that bone and prostate tissues rely on distinct path-region tokens, indicating that the decoder's routing is tissue-dependent.

Fig.~\ref{fig:physics}c shows the decoder's TA attention share, which varies smoothly across the image: different spatial locations draw on different paths, reflecting position-dependent routing learned without explicit supervision. 


\noindent\textbf{3.4 Ablation Study.}
We conduct an ablation study to demonstrate the importance of attention stages in MUX-USCT, as shown in Tab.~\ref{tab:ablation}.
From the results, removing transducer attention in encoder barely affects clean MAE ($+$0.78\,m/s) but sharply increases noise sensitivity (2.5$\times$ at $n{=}5$): the channel-weighting it provides is modest for clean signals yet critical when some channels degrade. Disabling decoder cross-attention raises clean MAE over fourfold to 30.93\,m/s, confirming per-pixel token routing as the core reconstruction mechanism. Together, the two attention stages are complementary: the encoder selects \emph{what} to listen to, and the decoder decides \emph{where} each signal contributes.

\begin{table}[t]
\centering
\caption{Ablation study. Noise columns: $n$/10 sources corrupted ($-$10\,dB).}
\label{tab:ablation}
\setlength{\tabcolsep}{4.5pt}
\begin{tabular}{@{}lcccc@{}}
\toprule
Configuration & Clean MAE & $n{=}3$ & $n{=}5$ & $n{=}7$ \\
\midrule
MUX-USCT (full) & \textbf{6.88} & \textbf{10.51} & \textbf{18.18} & \textbf{42.53} \\
\quad w/o transducer attention & 7.66 & 21.05 & 45.81 & 82.93 \\
\quad w/o decoder cross-attn & 30.93 & 26.93 & 33.21 & 55.07 \\
\bottomrule
\end{tabular}
\end{table}


\section{Conclusion}

MUX-USCT demonstrates that encoding the physical acquisition geometry into the network architecture improves both clean-data accuracy and robustness under diverse clinical degradations.
Although this paper uses a dual-probe prostate configuration, the design requires only that the probe layout and resulting path structure are known, a condition met by all synthetic aperture USCT systems.
The principle of learned selective routing of structured sensor channels can extend to other multi-channel sensing modalities (e.g., multi-coil MRI). 


\bibliographystyle{splncs04}
\bibliography{references}

\end{document}